\documentclass[12pt]{article}
\usepackage{amsmath,amssymb,amsfonts}
\usepackage{graphicx,psfrag,epsf}
\usepackage{enumerate}
\usepackage{natbib}
\usepackage{tikz}
\usepackage{pgf}
\usepackage{setspace}
\usepackage{hyperref}

\newtheorem{exm}{Example}

\begin{document}


\title{\bf Viewing Simpson's Paradox}
\author{Priyantha Wijayatunga \footnote{Department of Statistics, Ume\r{a} School of Business and Economics, Ume\r{a} University,  Ume\r{a} SE-90187, Sweden. Email: \emph{priyantha.wijayatunga@stat.umu.se}.}} 

\maketitle


\begin{abstract}
Well known Simpson's paradox is puzzling and surprising for many, especially for the empirical researchers and users of statistics. However there is no surprise as far as mathematical details are concerned. A lot more is written about the paradox but most of them are beyond the grasp of such users. This short article is about explaining the phenomenon in an easy way to grasp using simple algebra and geometry. The mathematical conditions under which the paradox can occur are made explicit and a simple geometrical illustrations is used to describe it. We consider the reversal of the association between two binary variables, say, $X$  and $Y$ by a third binary variable, say, $Z$. We show that it is always possible to define $Z$ algebraically for non-extreme dependence between $X$ and $Y$, therefore occurrence of the paradox depends on identifying it with a practical meaning for it in a given  context of interest, that is up to the subject domain expert.  And finally  we discuss the paradox in predictive contexts since in literature it is argued that the paradox is resolved using causal reasoning. 
\end{abstract}

\noindent%

\newpage
\section{Introduction} \label{sec:intro}

Simpson's paradox that was discussed originally in \cite{YG03} and later in \cite{SE51} but named in \cite{BC72} is a situation where we see that two random variables are positively (negatively) correlated but at the same time negatively (positively) correlated when given each value of a third variable. The paradox is found in many occasions in social science, epidemiological, economics, etc. applications. Nevertheless it is considered to be a puzzling and surprising phenomenon because of its contradictory conclusions when some interpretations of probabilities are used; for example, when causal interpretations are given to observed probabilities. Therefore it has received considerable attention in philosophical literature (see \cite{OR85}, \cite{BP10} and references therein) and in social science contexts as well. One famous example is about alleged  sex discrimination in University of California Berkeley graduate college admission discussed in \cite{BH75} where empirical data show that there is an overall  higher rate of admission for the male applicants but when the rates are considered academic department-wise there is a slight bias for the female students.  Further investigation reveals that the female applicants tend to apply more competitive departments where there are higher rates of rejection. However, if the smallest of  department-wise admission rates for the females  was greater than the largest of the department-wise admission rates for the males then the paradox would never have happened, whether or not the females  tended to apply for more competitive departments. That is, in the observed context if it were the case that all departments were favoring females highly (in that way) the application pattern of the sex groups could not have shown a different scenario than that when two groups are taken together. So, there is a numerically necessary condition for the paradox.   

Following is  another example  (data are adjusted and adapted from \cite{PJ09}). Note that here we avoid any small sample problems; for example, one can think that all the count are multiples of a large number such as $10 000$ so that every count  is large.  

\begin{exm}
Consider following table of counts that are obtained from an observed sample of individuals both males and females, who either had or had not taken a certain treatment for a certain disease where recovery from the diseases is also reported.
\begin{table}[h]
\caption{Numeric counts of recovery.  \label{tab:ex1}}
\begin{center}
\begin{tabular}{r|r|rr|r|rr|r|rrr}
\hline
                               & \multicolumn{2}{ c }{ Male ($Z=1$) } &       & \multicolumn{2}{ c }{ Female ($Z=0$) } &         & \multicolumn{2}{ c }{ Both }  &   \\
                               & \multicolumn{2}{ c }{ Recovery ($X$) } &      & \multicolumn{2}{ c }{ Recovery ($X$) } &       & \multicolumn{2}{ c }{ Recovery ($X$) } &    \\
                                     & Yes($1$) & No($0$)   &   & Yes($1$) & No($0$)                             &     & Yes($1$) & No($0$) & \\ \hline
Treatment ($Y$)      $1$  & $7$      & $3$     & & $9$  & $21$                                              &    & $16$  & $24$ &  \\
                               $0$  & $18$       & $12$     & & $2$ & $8$                                              &   & $20$  & $20$  & \\
\hline
\end{tabular}
\end{center}
\end{table}
The counts show that the treatment was effective for both the males and the females separately since its  recovery probabilities for the males and the females are greater than those of the non-treatment respectively;
\begin{eqnarray*}
\frac{7}{7+3} =0.7 > \frac{18}{18+12}=0.6   \qquad \textrm{and} \qquad  \frac{9}{9+21} =0.3 > \frac{2}{2+8} =0.2.
\end{eqnarray*}
However when we aggregate the data, i.e., pool the counts for the males and the females together, then we see that the treatment is not effective for the individuals anymore because then the recovery rate of the treatment is smaller than that of the non-treatment, i.e., 
\begin{eqnarray*} 
\frac{16}{16+24}=0.4 < \frac{20}{20+20} =0.5.
\end{eqnarray*}
\end{exm}
Clearly above interpretation of relative frequencies (probabilities) says an impossibility.  Of course numerically sum of the recovery probabilities (rates) of the treatment for the males and the females  are larger than that of the  non-treatment,  i.e., $\frac{7}{7+3} + \frac{9}{9+21} > \frac{18}{18+12} +\frac{2}{2+8}$ since the recovery probability of the treatment for the male is greater than that of the non-treatment and similarly for the females. In fact these sums  have no reasonable interpretation but one may tend to believe that the treatment group should always have a higher chance of recovery collectively perhaps due to the fact that it corresponds to the bigger summands, therefore the sum.  But the recovery probability of the treatment when the males and the females are taken together as a single group happens to be a certain weighted average of those separately for the males and the females and similarly for non-treatment. Write them as  
\begin{align*}
 \frac{7+3}{16+24} \cdot  \frac{7}{7+3}   &+  \frac{9+21}{16+24} \cdot \frac{9}{9+21} =\frac{16}{16+24}=0.4  \\
& <  \frac{18+12}{20+20} \cdot \frac{18}{18+12}  + \frac{2+8}{20+20} \cdot \frac{2}{2+8} = \frac{20}{20+20} =0.5 \\
u  \frac{7}{7+3} + (1-u) \frac{9}{9+21} =0.4  &< v \frac{18}{18+12}  + (1-v) \frac{2}{2+8} =0.5  \\
\end{align*}
where $u=\frac{7+3}{16+24}=0.25$ and $v=\frac{18+12}{20+20}=0.75$. Even if the sum of two probabilities  is greater than that of their complementary probabilities,  a certain (but desired) weighted average of the former two is not always  greater than that of the latter two with different (but desired) weights where weights sum to $1$ in each case. The key factor is that what the two sets of weights are.  As you have seen, in this case the two sets of weights are $\{u,1-u \}$ (for the recovery probabilities of the  treatment) and $\{v,1-v \}$ (for the recovery probabilities of the non-treatment). For some $u$ and $v$ that are implied by the counts in the Table \ref{tab:ex1} we can see that the recovery probability of the treatment for the individuals (when the males and the females taken together) can be smaller than that of the non-treatment. Note that it is not required that we have $u+v=1$, though here it is a coincidence. Furthermore $u$ is the probability of being a male in the treatment group and $v$ is that in the non-treatment group, so they indicate how variables sex and taking treatment are dependent. We see that for the paradox to happen it is necessary that $u$ is 'sufficiently' smaller than $v$, written as "$u<<v$". It means that the dependence between the two variables is "strong". But as in the previous example it is necessary that  the smallest of sex-wise  recovery probabilities of the treatment ($0.3$) is not greater than the largest of  sex-wise recovery probabilities of the non-treatment ($0.6$), i.e, the real value interval created by the recovery probabilities of the treatment for the males and the females (which is $[0.3,0.7]$) and that created by those of the non-treatment (which is $[0.2,0.6]$) should overlap, thus making an interval of probability values, otherwise the paradox can not occur.  

Whether or not one may have incorrect expectations about the rates calculated for the pooled group from those of its constituent groups, i.e., irrespective of numerical facts,  the above interpretation of the probabilities is clearly a paradox. This may be the reason that   it is stated in \cite{PJ09} that the paradox can be resolved by considering  causality, probabilistic causality to be more precise. Here the females tend to take treatment more often than the males do. This is taken to be a causal relation. While it is possible to have a positive association between a treatment and its outcome (recovery of from a disease) for both men and women, and a reversed association between treatment and recovery when the data are amalgamated, it is not possible for treatments to be causally effective for men and women, but not effective for people. 

Note that as one of the anonymous referees pointed out the Simpson's paradox can be explained by so called mediant fractions\footnote{for example, see \url{http://www.mathteacherctk.com/blog/2011/02/mediant-fractions-and-simpsons-paradox/}} (see also \cite{RJ2006}). However we do not wish to discuss the topic using them here.  So, let us discuss the paradox seen in the above example little more formally. We consider the simple case of the paradox with three binary variables; the reversal of the marginal association between $X$  and $Y$ by a third  variable $Z$. Let the set of possible values of $X$ be $\{x,x'\}$ where $x$ denotes the success and $x'$ denotes the failure of an event of interest, and similarly for $Y$ and $Z$. One can get these results for a more general case where $Z$ is multi-nary. One instance of the Yule-Simpson's paradox says that the following relationships between conditional probabilities are possible.
\begin{eqnarray}
p(x \vert y,z) \geq p(x \vert y',z)   \label{pxy1z} \\
p(x \vert y,z') \geq p(x \vert y',z') \label{pxy1z1}
\end{eqnarray}
but at the same time
\begin{eqnarray}
p(x \vert y) < p(x \vert y')    \label{px1y}
\end{eqnarray}
Equivalently 
\begin{eqnarray}
p(x,y \vert z) \geq p(x \vert z)p(y \vert z)    \label{cdxy_z} \\
p(x,y \vert z') \geq p(x \vert z')p(y \vert z')  \label{cdxy_z1}
\end{eqnarray}
but at the same time
\begin{eqnarray}
p(x,y) < p(x )p(y)  \label{pxy_xy}
\end{eqnarray}
Conditional dependences between $X$ and $Y$ in the two cases of when it is given that $Z=z$ and $Z=z'$ is non-negative but marginal dependence between them is negative. Note that the other occurrence of the paradox is similar. It can be obtained by, for example, interchanging the naming of the success and the  failure of the event related to the  variable $X$.

Empirical researcher may find that this phenomenon is surprising but the algebra behind this reversal is not. The paradox is often a numerical possibility. The main idea about this short article is to make the mathematical details of the paradox explicit (as agreed by one of the anonymous referees) and to show a useful graphical representation on how the paradox can occur.  This explanation of the paradox is particularly useful for the empirical researchers for explaining the context and also for students of statistics.  When $X$ and $Y$ are dependent through a non-extreme conditional probability (i.e., $0 < P(X \vert Y ) <1$), using the  geometric figure it is easy to see that it is always possible to define another third binary variable, say, $Z$ that induces reversed dependences between $X$ and $Y$ at each value of $Z$. Note that a variable is a criterion that  makes a partition in the collection of subject (the observed sample, in this case). Therefore, there can be  many variables that can be defined on a given sample.  However such  $Z$ may only be an algebraic and hypothetical, if not subjective, variable unless it is not possible to give a real meaning to it, for example, finding $Z$ as a hidden variable. So, we argue that the occurrence of the paradox in this case is completely dependent on finding such $Z$ that has a real subject domain meaning. So, it is up to the subject domain expert to decide if $Z$ makes sense. Conditional probabilities of occurrence of the paradox is discussed in \cite{PP2009}.   However, our argument is that the possibility of the paradox is domain dependent, i.e, finding meaningful $Z$ in the context. It has no relation whatsoever to the probability $P(X \vert Y)$ which is positive. Note that often researchers concern about if there exits such $Z$. In fact, algebraically there can be infinitely many $Z$ for non-extreme  $P(X\vert Y)$. And on the other hand when we know the conditional probabilities $P(X \vert Y, Z)$ then we can find sufficient conditions that are simple to avoid the occurrence of the paradox.  And one can extend the above discussion for the case of multinary $Z$ where the geometrical illustration is not easy. 

Though it is reasonable to believe that for the occurrence of the paradox there should be some causal relations between $X$ and $Z$  or $Y$ and $Z$, one may find other situations where there are no such obvious causal relations. We discuss one example taken from current philosophical literature for the latter, that is a phenomenon by design. We also discuss another example of the paradox in a predictive context. In some early literature it was discussed how to proceed with a new case in the paradoxical context, i.e., how a new case should be treated; either using marginal relationship between $X$ and $Y$ or conditional relationships between $X$ and $Y$ given $Z$, that are shown to us in the observed data. Often such discussions on resolving the paradox is done by using some criteria and so-called causal calculus found in graphical modeling theory (see \cite{PJ09}). We avoid those discussions here.

\section{Algebraic and Graphical Explanation}
\label{sec:alge}

Now let us see how the paradox  can happen with algebra of probabilities. Suppose that we have above instance of the paradox  and let us multiply the inequalities (\ref{cdxy_z})
and (\ref{cdxy_z1}) with the inequalities $ p(z)  \geq (p(z))^2$ and $p(z') \geq (p(z'))^2$ respectively and then we get
\begin{eqnarray}
p(x,y,z) \geq p(x,z)p(y,z)     \\
p(x,y,z') \geq p(x,z')p(y,z')  
\end{eqnarray}
By adding them together it gives 
\begin{eqnarray}
p(x,y) \geq p(x,z)p(y,z)+ p(x,z')p(y,z')
\end{eqnarray}
But $p(x)p(y) = (p(x,z)+p(x,z'))(p(y,z)+p(y,z'))$ implies that 
\begin{eqnarray}
p(x,y) \geq p(x)p(y) - (p(x,z)p(y,z')+ p(x,z')p(y,z))   \label{pxy}
\end{eqnarray}
Since $(p(x,z)p(y,z')+ p(x,z')p(y,z))$ is non-negative dropping it from expression (\ref{pxy}) may not  preserve the inequality in the current form so sometimes we get the result $p(x,y) \leq p(x)p(y)$ which is the inequality (\ref{pxy_xy}). So, algebraically the paradox is  simple and can occur sometimes.

Furthermore, alternatively, when $p(z \vert y) \leq p(z \vert y')$ then it implies that $p(z' \vert y) > p(z' \vert y')$. By multiplying the inequalities (\ref{pxy1z}) and (\ref{pxy1z1}) with these restrictions respectively we get 
\begin{eqnarray}
p(x,z \vert y) \lesseqgtr  p(x,z \vert y') \\
p(x,z' \vert y) \geq p(x,z' \vert y') 
\end{eqnarray}
And now if you add them together the we get
\begin{eqnarray}
p(x \vert y) \lesseqgtr p(x \vert y').
\end{eqnarray}
This implies that either of inequalities can occur where the paradox happens if it results in $p(x \vert y) < p(x \vert y') $. Note that the other case of $p(z \vert y) > p(z \vert y')$ (which implies that $p(z' \vert y) \leq p(z' \vert y')$) results in the same conclusion. So it is clear that for the occurrence of  the paradox it necessary that $Y$ and $Z$ are 'sufficiently' dependent. 

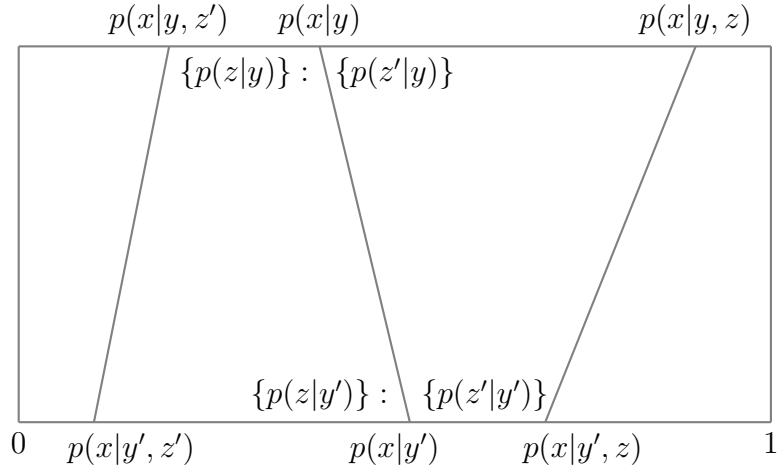
\begin{figure}
\begin{center}
\caption{An occurrence of the Simpson's paradox: for probability $p$, indication of $\{p\}$:$\{1-p\}$ means that the lengths of two line segments on which $\{p\}$ and $\{1-p\}$ appear are according to the ratio $p:1-p$. \label{fig:fig1}}
\begin{tikzpicture}[scale=10]
    \draw [thick, gray, -] (0,0) -- (0,0.5);      

    \draw [thick, gray, -] (0,0) -- (1,0);      


    \node [below] at (1,0) {$1$};               
    \node [below] at (0,0) {$0$};               

\draw [thick, gray, -] (0,0.5) -- (1,0.5);
\draw [thick, gray, -] (1,0) -- (1,0.5);
\draw [thick, gray, -] (0.1,0) -- (0.2,0.5);
\node [below] at (0.15,0) {$p(x \vert y',z')$};
\node [below] at (0.75,0) {$p(x \vert y',z)$};

\draw [thick, gray, -] (0.7,0) -- (0.9,0.5);
\node [above] at (0.2,0.5) {$p(x \vert y,z')$};
\node [above] at (0.9,0.5) {$p(x \vert y,z)$};

\draw [thick, gray, -] (0.52,0) -- (0.4,0.5);
\node [below] at (0.5,0) {$p(x \vert y')$};
\node [above] at (0.4,0.5) {$p(x \vert y)$};

\node [above] at (0.4,0) {$ \{ p(z \vert y') \}: $};
\node [above] at (0.62,0) {$ \{p(z' \vert y') \} $};

\node [below] at (0.3,0.5) {$ \{ p(z \vert y) \}: $};
\node [below] at (0.5,0.5) {$ \{p(z' \vert y) \} $};
\end{tikzpicture}
\end{center}
\end{figure}

Let us consider the case given by the expressions (\ref{pxy1z}),  (\ref{pxy1z1}), say, Case 1, where the paradox is when the expression (\ref{px1y}) is true. Then $p(x \vert y) < p(x \vert y') $ gives $ p(z' \vert y)p(x \vert y,z') + p(z \vert y)p(x \vert y,z) < p(z' \vert y')p(x \vert y',z') + p(z \vert y')p(x \vert y',z) $. The left hand side of the inequality is the weighted average of the conditional probabilities $p(x \vert y,z')$ and  $p(x \vert y,z)$ where weights sum up to $1$ (i.e., $p(z' \vert y) + p(z \vert y)=1$). And similarly for the right-hand side of the inequality. Since, when the weights are positive then the  weighted average of two numbers is contained in the interval whose end points are the two numbers it is easy to see that the two intervals corresponding to these four conditional probabilities should overlap  each other for the possibility of the occurrence of the paradox. See the Figure~\ref{fig:fig1} for this case of the relationships among the conditional probabilities where the two intervals are marked on two parallel horizontal lines.  It is necessary but not sufficient that we have $ Min \{p(x \vert y, z), p(x \vert y, z') \} < Max \{ p(x \vert y', z), p(x \vert y',z')\}$ for the paradox to occur. Therefore sufficient condition for non-occurrence of the paradox is that $ Min \{p(x \vert y, z), p(x \vert y, z') \} \geq Max \{ p(x \vert y', z), p(x \vert y',z')\}$. However under the necessary condition for the paradox, not having certain dependence between $Y$ and $Z$ avoids the paradox (as we seen in the example). In the following we assume that the necessary condition holds. 

It is simple yet important to note that the value $p(x \vert y)$ dissects  positive length $p(x \vert y,z)-p(x \vert y,z')$ according to the ratio $p(z \vert y):p(z' \vert y)$;
\begin{align*}
\{p(z \vert y)+p(z' \vert y)\}p(x \vert y)&=p(z' \vert y)p(x \vert y,z') + p(z \vert y)p(x \vert y,z) \\
p(z \vert y) \{p(x \vert y,z)-p(x \vert y) \} &= p(z' \vert y) \{p(x \vert y)-p(x \vert y,z') \} \\
\frac{p(x \vert y)-p(x \vert y,z')}{p(x \vert y,z)-p(x \vert y)} &=\frac{p(z \vert y)}{p(z' \vert y)}
\end{align*}
And similarly the value $p(x \vert y')$ dissects  positive length $p(x \vert y',z)-p(x \vert y',z')$ according to the ratio $p(z \vert y'):p(z' \vert y')$. In the Figure~\ref{fig:fig1}  those ratios are marked  with braces. And from the above expression of weighted averages;
\begin{align*}
p(x \vert y,z') & + p(z \vert y) \big\{ p(x \vert y,z) - p(x \vert y,z') \big\} \\
& < p(x \vert y',z') +p(z \vert y') \big\{ p(x \vert y',z) - p(x \vert y',z') \big\}  \\
p(x \vert y,z') & + \Bigg\{ \frac{p(z \vert y)}{p(z \vert y)+p(z' \vert y)} \Bigg\} \big\{ p(x \vert y,z) - p(x \vert y,z') \big\} \\
& < p(x \vert y',z') +\Bigg\{\frac{p(z \vert y')}{p(z \vert y')+p(z' \vert y')}\Bigg\} \big\{ p(x \vert y',z) - p(x \vert y',z') \big\} 
\end{align*}
It is clear that when numerical $p(z \vert y)$  fraction of the length the $p(x \vert y,z) -p(x \vert y,z')$  added to  $p(x \vert y,z')$ is smaller than numerical  $p(z \vert y')$ fraction of the length $p(x \vert y',z) -p(x \vert y',z')$  added to  $p(x \vert y',z')$ the paradox can occur. In other words, the occurrence of the paradox depends on the conditional probability $P(Z  \vert Y )$, i.e., the dependence between $Y$ and $Z$ (and equivalently $P(Z  \vert X )$) given that  the necessary condition for the paradox is satisfied. That is, in this case for occurrence of the paradox $p(z \vert y)$ should be sufficiently smaller than $p(z \vert y')$, written as $p(z \vert y)<<p(z \vert y')$ as previously. 

Note that other three cases, namely, Case 2:  $p(x \vert y,z)< p(x \vert y,z')$ and $p(x \vert y',z)< p(x \vert y',z')$, Case 3: $p(x \vert y,z)< p(x \vert y,z')$ and $p(x \vert y',z')< p(x \vert y',z)$ and Case 4: $p(x \vert y,z')< p(x \vert y,z)$ and $p(x \vert y',z)< p(x \vert y',z')$ can be treated similarly. And furthermore, for a more general case where $Z$ taking more than two values one can easily get all the above algebraic relations. But it may be difficult to mark the corresponding conditional probability ratios in the geometric figure for  such cases  due to overlapping of distances.

\section{Causal Confounder or Reversal Attribute} \label{sec:cau}

Now we turn into some philosophical literature about the paradox. Recently, in \cite{BP10} it is argued that there can be completely non-causal cases of the paradox though it is shown in \cite{PJ09} and \cite{AO08} that it can be resolved by considering causality, implying that solution to the paradox needs causal explanation. Furthermore, in \cite{AO08} it is argued that the paradox is a problem of covariate selection and adjustment (when to control for or not) in causal analysis of non-experimental data. Note that these types of discussions are important for statisticians too. Following their arguments one can create a case such as follows. Suppose that we have two packs of cards. Let the variable pack refers to variable sex ($Z$) in our Example~\ref{tab:ex1} such that value 'pack 1' refers 'male' and value 'pack 2' refers 'female'. On each card a circle is drawn either big or small using either of two colors of red and blue. Let the variable size refers taking the treatment ($Y$) such that the value 'big' refers to 'treatment' and the value 'small' refers to 'non-treatment'. And the variable color refers to recovery ($X$) such that value 'red' refers to 'recover' and value 'blue' refers to 'not recover'. Assume that we randomly draw cards one at a time with replacement from the packs. Then the probability of getting a red card when it is with a big circle is greater than that of when it is with a small circle for each of the packs separately, i.e., $P$(red $\vert$ big, pack 1) = 0.7 $ > P$(red $\vert$ small, pack 1) = 0.6 and $P$(red $\vert$ big, pack 2) = 0.3 $ > P$(red $\vert$ small, pack 2) = 0.20. But if we pool the cards of the two packs together then the probability of getting a red card when it is with a big circle on it is smaller than that of  when it is with a small circle on it, i.e., $P$(red $\vert$ big) = 0.4  $ < P$(red $\vert$ small) = 0.50, so the paradox seems to occur.

Arguably it is not natural that size of the circle  causes it  to be red or blue and vice versa but there is a dependence between size and the color of the circle in this context and similarly for pack and color and pack and size of the circle. So, naturally and simply one can argue against causality in this context.  However causality is a difficult concept to discuss here. But often such discussions are inevitable in observational data analysis.  The context can be entirely a predictive one, however it is by design. And the paradox may not be resolved as long as it is not given how the card is taken -whether it is selected randomly from either of the packs or from the pooled set of cards.  When it is given how it is taken the paradox is immediately resolved. That is, in the case of missing information on how the card is selected the  paradox seems to exist. But one can ideally calculate  desired probabilities, for example, by assuming some probability for pooling two card packs such as $0.5$ (therefore, not pooling has probability $0.5$) and similarly for selecting a pack when card packs are not pooled (along with all other conditional probabilities given). So, the paradox can be resolved in this predictive context by looking for any missing information in the context, i.e., finding it exactly or assuming some probability on it. Though this is not a complete proof of solving the paradox in predictive contexts, we argue that the paradox seems to exist as long as we face with missing information. If one is capable of assuming correct probabilities for the missing details of the context or find them exactly  then there is no paradox. 

For clarity suppose that it is equally likely that the card is selected from one of the two packs or from the pooled pack and furthermore, in the former case it is also equally likely that it is selected from either of the two packs. And let these possibilities are connected with a random variable, say,  $T$ where $P(T=1)=0.25$, $P(T=2)=0.25$ and $P(T=3)=0.5$. Then, if the selected card has big circle on it then the probability that it is a red circle is,
\begin{eqnarray*}
p(\textrm{red} \vert \textrm{big}) &=& \sum_t p(\textrm{red}, T=t \vert \textrm{big}) =\sum_t p(\textrm{red} \vert \textrm{big},T=t) p(T=t \vert \textrm{big}) \\
                           &=& \sum_t p(\textrm{red} \vert \textrm{big},T=t) p(T=t ) \\
                       &=&  p(\textrm{red} \vert \textrm{big}, \textrm{pack} 1) \times 0.25 +  p(\textrm{red} \vert \textrm{big}, \textrm{pack} 2) \times 0.25 \\
                  & & \qquad + p(\textrm{red} \vert \textrm{big}, \textrm{pooled pack} ) \times 0.5  \\
                  &=& 0.7 \times 0.25 + 0.3 \times 0.25 + 0.4 \times 0.5 = 0.45
\end{eqnarray*}
Likewise any other desired probability can also be calculated.  Note that here $p(\textrm{red} \vert \textrm{big})$ is calculated for the case where the action of selection has been taken place (i.e., considering all possible values for $T$), so $T$ is assumed to be independent of anything. Furthermore, such predictive contexts have no relevance to causal contexts where there are possibilities of assigning values for $Y$ for the new subject.

Let us consider another predictive context as follows. In a certain newly created city, there are people who are either white or black coming from either northern part or southern part of the country and speaking either of two languages, say, $A$ and $B$ as their mother tongue.  Let we have taken a random sample of people from the city population and assume that the corresponding counts are as in the Table~\ref{tab:ex1}. Here we have taken skin color as the variable treatment ($Y$) where 'white' refers to 'treatment' and 'black' refers to 'non-treatment', the mother tongue refers to the variable recovery ($X$) where the 'language A' refers to 'recover' and the 'language B' refers to 'not recover'  and the region that they are coming from refers to the variable sex ($Z$) where  'male' refers  to 'north' and 'female' refers to 'south'. Imagine that all of the city population are new arrivals and therefore any selected person  can not be intervened to have any desired value for any of those three variables. That is, for any given person from the city population we can only do predictions on these three variables. So, it is completely a predictive context.  Now for a randomly selected person, say, a white  person, whether or not the probability that his or her mother tongue is the language A is greater than that of the language B depends on how the person is selected; either from all the people who are originally coming from same region or  from all the people living in the city irrespective of their original region of coming from. That is, we need to know the selection mechanism to answer the predictive problem or otherwise we need to assume a certain probability on the selection mechanism, as in the case of card example above. 

Now consider the following predictive case. Let some one from city population comes  forward and challenges us to predict  his or her mother tongue, perhaps knowing about paradoxical conclusion in our random sample of data. Then it can be a hard time for us to know about the person's self-selection as well as to assume some probability on its mechanism. However it is always possible for us to assume  a probability  as required and do the prediction as in the case of card packs.  

In both predictive contexts above, the paradox is resolved immediately when we know about the selection mechanism of the new subject, i.e., no missing information. This is similar to the case of knowing about whether or not the variable $Z$ is causally affecting the variable $Y$ in the causal contexts. But the causal contexts have other possibilities even within this case. In the predictive contexts we can do predictions even when how the selection is done is not known.  But in the causal context of the paradox, we can not perform similar tasks as we are required to assign a value to $Y$; we can not intervene the new subject more than once. For this reason, the paradox has more significant aspects in causal contexts than those in predictive contexts. If someone finds paradox is surprising  in a predictive context, it is due to difficulties in understanding how ratios for the pooled group are made from those of its subgroups.      

\section{Conclusion}
Here we have given an explicit mathematical explanation of the Simpson's paradox  using simple algebra and a geometric figure. These details help the empirical researchers and the students of statistics to understand the nature of the paradox. We have seen that it is always possible to define a third variable, say,  $Z$ algebraically for any non-extreme dependence between two other variables, say, $X$ and $Y$, (that is, when $P(X \vert Y)>0$) so that the paradox occurs. So, for a given such context meaningfulness of the paradox is dependent on identifying $Z$ as a hidden variable with a real practical meaning, that is up to the subject domain expert. And on the other hand it is easy to see the algebraic conditions to avoid the paradox in the case of obtaining $P(X \vert Y)$ when $P(X \vert Y,Z)$ is known.

And finally  we have discussed  some predictive contexts where the paradox can occur.  It occurs when we do not have sufficient information on the context. However, in literature it is generally accepted that paradox can be resolved with causal knowledge. Of course, there exit causal contexts of the paradox, that requires causal knowledge to resolve it. Causal contexts are harder to resolve than predictive contexts where one can assume some probabilities to do predictions so that overall predictive accuracy is acceptable. 

\section{Acknowledgments}
The author gratefully acknowledges the financial support of the Swedish Research Council through the Swedish Initiative for Microdata Research in the Medical and Social Sciences (SIMSAM)  and the Swedish Research Council for Health, Working Life and Welfare (FORTE).

{}

\end{document}